\documentclass[12pt]{article} 
\usepackage{latexsym} 
\def\a{\alpha} 
\def\b{\beta} 
\def\g{\gamma}  
\def\d{\delta}   
\def\h{\eta}
\def\l{\lambda}

\def\m{\mu}  
\def\n{\nu}  
  
\def\o{\omega}
\def\z{\zeta}

\def\eps{\epsilon}
\def\x{\xi}

\def\pa{\partial}

\def\be{\begin{equation}}  
\def\ee{\end{equation}}  
\def\beq{\begin{eqnarray}}  
\def\eeq{\end{eqnarray}}

\def\cc{{\cal C}}

\def\cl{{\cal L}}

\newcommand{\bqn}{\begin{eqnarray}}\newcommand{\eqn}{\end{eqnarray}}
\newtheorem{theorem}{Theorem}[subsubsection]

\begin{document} 
\begin{titlepage} 
\begin{flushright}
ULB-TH/01-32\\
\end{flushright}
\vskip .5cm
\begin{centering} 
 
{\huge {\bf A note on the uniqueness of  $D=4, N=1$ Supergravity}} 
 
\vspace{3cm} 
 
{\Large Nicolas Boulanger$^{a,}$\footnote{"Chercheur F.R.I.A.", Belgium} and  
Mboyo Esole $^{a}$ } \\ 
\vspace{.4cm} 
$^a$ Physique Th\'eorique et Math\'ematique,  Universit\'e Libre 
de Bruxelles,  C.P. 231, B-1050, Bruxelles, Belgium      \\ 
\vspace{.2cm}  
 
{\tt nboulang@ulb.ac.be, mesole@ulb.ac.be}

\vspace{2cm} 
 
\end{centering}  
\begin{abstract} 
We investigate in 4 spacetime dimensions, all the consistent deformations
 of the lagrangian ${\cal L}_2+{\cal L}_{\frac{3}{2}}$,
 which is the sum of the Pauli-Fierz lagrangian  ${\cal L}_2$ for
 a free massless spin 2 field and the Rarita-Schwinger
 lagrangian  ${\cal L}_{\frac{3}{2}}$ for 
a free massless spin 3/2 field.
\par Using BRST cohomogical techniques, we show, under the assumptions of
 locality, Poincar\'e invariance,
 conservation of the number of gauge symmetries and the number 
of derivatives on each fields, that N=1 D=4 supergravity is the only
consistent interaction between a massless spin 2 and a massless spin 3/2 
field. 
We do not assume general covariance. 
This follows automatically, as does supersymmetry invariance.
Various cohomologies related to conservations 
laws are also given.
\end{abstract}    
\vfill           
\end{titlepage}  

\section{Introduction} 
\setcounter{equation}{0} 
\setcounter{theorem}{0} 
\setcounter{lemma}{0} 

It is well appreciated that general relativity is the unique way to
consistently deform the Pauli-Fierz action $\int\mathcal{L}_2$ for a free
massless spin-2 field under the assumption of locality, Poincar\'e
invariance, preservation of the number of gauge symmetries and the number of
derivatives in $\cl_2$ \cite{Gupta, Kraichnan, Feynman,Weinberg,Ogiev,Wyss,
Deser1,Fronsdal,Berendsetal, Wald0}.
This has been reconfirmed recently in \cite{graviton} using 
BRST-cohomological techniques based on the antifield
formalism, where multi-graviton theories were also included.\newline

Supersymmetry seems to be crucial in the attempts to reconcile quantum
mechanics and gravitation. It is then natural to consider the Rarita-Schwinger
 action $\int \mathcal{L}_{\frac{3}{2}}$
for a free massless spin 3/2 field, which describes the gravitino, the
supersymmetric partner of the graviton and to subsequently analyse all the 
consistent deformations of the free action $I_0=\int \cl_2+\cl_{\frac{3}{2}}$.
\newline

In this paper we show that, under the assumptions of locality,
Poncar\'e invariance, conservation of the number of gauge symmetries and the
number of derivatives on each fields,
\newline

(i) the only consistent deformation of the
lagrangian $\mathcal{L}_{\frac{3}{2}}$ is given to first order in the
coupling constant $m$, by a mass term $\mathcal{L}_{M}$ which is obstructed
at second order (the appearance of this mass term is a well known property
of the full nonlinear N=1 D=4 supergravity \cite{cosmological_constant} ).

(ii) the most general deformation of the lagrangian 
$\cl_2+\cl_{\frac{3}{2}}$ to first order is given by 
\be
\cl=\cl_2+\cl_{\frac{3}{2}}+g\cl_{E}+\Lambda
\cl_{C}+m\cl_{M}+\alpha\cl_{Int} 
\nonumber 
\ee
where $\mathcal{L}_{E}$ is the cubic vertex of Einstein-Hilbert lagrangian
(containing two derivatives :  ``$\pa\pa h h h$''), 
$\mathcal{L}_{C}=-2h^{\m\n}\h_{\m\n}$ is
the first order deformation of the Pauli-Fierz action which corresponds to
the cosmological term and $\mathcal{L}_{Int}$ is the unique consistent
deformation to first order in the coupling constant $\alpha $ involving the
two fields (graviton and gravitino) simultaneously. This last term involves 
the spin
connection to first order, {\it{i.e.}} converts ordinary derivatives of  
the (vector-)spinor
field into covariant derivatives. The introduction of general
covariance is not assumed and follows automatically.
This strengthens previous results 
\cite{consistent_superg,progress_supergr,properties_supergravity,supergravity}. 

(iii) consistency to second order in the coupling constants requires relations
between the constants : $g=4\alpha$ and $3m^2=\alpha\Lambda$ 
(see \cite{broken}).
This leaves two independent coupling constants.  As it is known that
supergravity $N=1$ $D=4$ (with a possible cosmological term) is a consistent
deformation of $\mathcal{L}_2+\mathcal{L}_{\frac{3}{2}}$ to all orders in the
coupling constants, we conclude that it is in fact the
unique consistent interaction between a spin-2 field and a spin-3/2 field under
the assumptions made above.
\newline

In fact, our paper can be viewed to some extend as a cohomological version of 
the Noether approach to supergravity developed in the pioneering paper
\cite{Deser:zb}. We should point out, however, that we do not assume a priori
the Noether form "currents times gauge fields" for the coupling : this 
follows automatically from our requirements as the sole consistent possibility.

We would like to stress that $\Lambda \cl_C$ is a consistent first order 
deformation \cite{graviton} of the Pauli-Fierz action  
$S^{PF}[h_{\m\n}]=\int \cl_2$,  
and that the presence of this cosmological term is not in conflict with 
the fact that we
wrote the metric as a perturbation around the flat Minkowskian metric
$g_{\m\n}=\h_{\m\n}+g h_{\m\n}$, $g$ being the deformation parameter.
The cosmological term can arise at order $g$ or higher in the deformation 
because it is compatible with the gauge symmetries.
For an analysis where the full Einstein action is derived 
by consistent self-coupling requirements from the linear graviton 
action in an a priori arbitrary background geometry where the cosmological
constant is present already at order zero, see \cite{Deser:uk}.
It would be interesting to perform the same analysis using the BRST 
cohomogical techniques.

The use of the antifield formalism streamlines the result and systematizes
the search for all possible consistent interactions
 of the free $I_0=\int \cl_{2} + \cl_{\frac{3}{2}}$ theory. 
We recover in a unified 
and esthetic way famous
results on $D=4$ $N=1$ supergravity (relation between the coupling constants,
appearance of the mass term with an abelian algebra,...) with fewer
assumptions (we do not assume general covariance, it follows automatically,
 as does the supersymmetric invariance). We also compute various
cohomologies related to conservations laws. This will be useful  for 
the study of  different aspects of the full $D=4$, $N=1$ theory as well
as its extensions with more supersymmetries\footnote{
The BRST-cohomology of the full nonlinear supergravity $D=4$ $N=1$ has been 
studied in \cite{Brandt_supergravity}}, thanks to the
importance of BRST approach for renormalization and anomalies.

\subsection{Conventions}

We work in 4-dimensions Minkowski space-time with the metric 
$\eta _{\mu \nu }=diag(1,-1,-1,-1)$. $\epsilon ^{0 1 2 3
}=-\epsilon _{0123 }=1$, $\{\gamma_{\mu},\gamma_{\nu}\}=
2\eta_{\mu\nu}$, the
gamma matrices $\gamma _{\mu }$ are
all purely imaginary. We take the matrix $\gamma ^{0}$ 
to be antisymmetric and hermitian, $\gamma ^{i}$ $(i=1,2,3)$ are symmetric
and antihermitian. The Dirac conjugate of a spinor is $\bar{\psi}=\psi
^{\dagger }\gamma ^{0}$ and the Majorana conjugate is $\psi ^{c}=(\mathcal{%
C\psi )}^{\top }$. The charge conjugate matrix $\mathcal{C}$ 
such that $\cc\g_{\m}=-\g_{\m}^{\top }\cc$
is given in our
conventions by $-\gamma ^{0}$. The Majorana spinors are such that $\bar{\psi}%
=\psi ^{c}$ and are therefore real.

We define \footnote{The notation $[a_1\ldots a_n]$ and $(a_1\ldots a_n)$ 
means that we
consider the expression which is totally antisymmetric (resp. symmetric) in
all the indices $a_1\ldots a_n$ with the normalization factor $\frac{1}{n!}$
, {\it{i.e.}} $\gamma^{[\mu}\g^{\nu]}\equiv\frac{1}{2}(\gamma^{\mu}
\gamma^{\nu}-\gamma^{%
\nu}\gamma^{\mu})$} : $\gamma^{\mu\nu}=\gamma^{[\mu}\gamma^{\nu]} %
\mbox{\hspace{0.5cm}} \gamma^{\mu\nu\rho}=\gamma^{[\mu}\gamma^{\nu}\gamma^{%
\rho]}$.

\subsection{The Free Models}

The  Pauli-Fierz lagrangian \cite{Pauli_Fierz} is given by : 
\begin{eqnarray}  \label{PF}
\mathcal{L}_{2}=-\frac{1}{2}
\partial_{\mu}h_{\nu\rho}\partial^{\mu}h^{\nu\rho}
+\partial_{\mu}h^{\mu}_{\nu}\partial_{\rho}h^{\rho\nu}
-\partial_{\nu}h\partial_{\rho}h^{\rho\nu} +\frac{1}{2}\partial_{\mu}h%
\partial^{\mu}h
\end{eqnarray}
where $h_{\mu\nu}$ is a covariant symmetric tensor of rank 2. The action 
$S^{PF}=\int d^4x \cl_2$ is
invariant under the irreducible and abelian gauge transformations : 
\begin{equation}  \label{graviton_gauge}
\delta_{\eta}h_{\mu\nu}=\partial_{\mu}\eta_{\nu}+\partial_{\nu}\eta_{\mu}
\end{equation}
where $\eta_{\nu}$ is a 4-vector.

The Rarita-Schwinger lagrangian \cite{Rarita_Schwinger} is given by : 
\begin{equation}  \label{RS}
\mathcal{L}_{\frac{3}{2}} = -\frac{1}{2}\bar{%
\psi}_{\alpha}\gamma^{\alpha\mu\nu}\partial_{\mu}\psi_{\nu}
\end{equation}
where $\psi_{\mu}$ is a fermionic Majorana spinor-vector. The action 
$S^{RS}=\int d^4x \cl_{\frac{3}{2}}$ is
invariant under the irreducible abelian gauge transformations : 
\begin{equation}  \label{gravitino_gauge}
\delta_{\epsilon}\psi_{\mu}= \partial_{\mu}\epsilon
\end{equation}
where the gauge parameter $\epsilon$ is a fermionic Majorana spinor.

\section{Cohomological reformulation}
\setcounter{equation}{0}  
\setcounter{theorem}{0}  
\setcounter{lemma}{0}  

\subsection{Differentials $\delta$, $\gamma$ and $s$}
\label{diff}

By following the general prescription of the antifield
formalism \cite{ Batalin_Vilkovisky, gauge_systems, rept_cohomology}, one
finds that the spectrum of fields, ghosts and 
their associated antifields is given by :

\begin{itemize}
\item[$\bullet$] the field $h_{\mu \nu }$ (the graviton);

\item[$\bullet$]  the field $\xi _{\nu }$, the ghost associated to the gauge
transformations (\ref{graviton_gauge});

\item[$\bullet$]  the antifield $h^{*\mu \nu }$ conjugated to the field 
$h_{\mu \nu }$;

\item[$\bullet$]  the antifield $\xi ^{*\nu }$ conjugated to the ghost 
$\xi_{\nu }$.

\item[$\bullet$] the field $\psi _{\mu }$ (the gravitino);

\item[$\bullet$]  the field $C$, the ghost associated with the gauge
symmetries (\ref{gravitino_gauge});

\item[$\bullet$]  the antifield $\bar{\psi}^{*\mu }$ conjugated to the field 
$\psi _{\mu }$;

\item[$\bullet$]  the antifield $\bar{C}^{*}$ conjugated to the ghost $C$.
\end{itemize}

We introduce the differential $\gamma$ which is the longitudinal derivative 
along the gauge orbits, and $\delta$ which is the 
Koszul-Tate differential related to the equations of motion.
A grading is associated to each of these differentials :
$\g$ increases by one unit the "pure ghost number" denoted {\it{puregh}}
while $\d$ increases the ``antighost number'' {\it{antigh}} by one unit.
The BRST-operator $s$ is simply the sum of the two differentials :
\be
s=\g +\d.
\ee
The ghost number {\it{gh}}, in turn, is defined by
\be
{\it{gh}}={\it{puregh}}-{\it{antigh}}.
\ee
The action of the differentials $\gamma$ and $\delta$  
on all the fields of the formalism is displayed in the following array which
indicates also the  pureghost number, antighost number, ghost 
number and grassmannian parity of the various fields : \vspace{5mm}

\begin{center}
\begin{tabular}{|c|c|c|c|c|c|c|}
\hline
Z & $\gamma(Z)$  & $\delta(Z)$  & $puregh(Z)$  & $antigh(Z)$  & $gh(Z)$  &  
Grassmannian parity \\ \hline
$h_{\mu\nu}$  & $2\partial_{(\mu}\xi_{\nu)}$  & $0$  &$0$  & $0$  &$0$ &$0$ \\ 
$\xi_{\mu} $ & $0$ & $0$ & $1$ & $0$ & $1$ & $1$ \\ 
$h^{*\mu\nu}$ & $0$ & $\frac{\delta^R }{\delta h{\mu\nu}}\mathcal{L}_2$ & $0$
& $1$ & $-1$ & $1$ \\ 
$\xi^{*\nu}$ & $0$ & $-2\partial_{\mu}h^{*\mu\nu}$ & $0$ & $2$ & $-2$ & $0$
\\ \hline
$\psi_{\mu}$ & $\partial_{\mu}C$ & $0$ & $0$ & $0$ & $0$ & $1$ \\ 
 $ C$ &$0$ & $0$ & $1$ & $0$ & $1$ & $0$ \\ 
$\bar{\psi}^{*\mu}$ & $0$& 
 $\frac{\delta^R }{\delta \psi_{\mu}}\mathcal{L}_{\frac{3}{2}}$ 
& $0$ & $1$ & $-1$ & $0$ \\ 
 $\bar{C}^{*}$ & $0$ & $\partial_{\mu}\bar{\psi}^{*\mu}$ & $0$ & $2$ & $-2$ & 
$1$  \\ \hline
\end{tabular}
\end{center}
 
\vspace{5mm} 

It is easy to check that :
\bqn
&\gamma^2=\delta^2=\gamma\delta+\delta\gamma=0,&
\nonumber \\ 
&s^2=0.&
\nonumber 
\eqn

\subsection{Consistent deformations and cohomology}

We analyse the problem of consistent deformation in the light of the master
equation formalism \cite{master_equation}. For a review, see \cite
{gauge_systems, rept_cohomology, consistent_interactions}.

The master equation formalism associates to a local action $I_0[\phi^i]$,
which is invariant under the gauge transformations \footnote{%
We use the De Witt's condensed notation : a summation over a repeated index
implies also an integration over spacetime variables.} 
\begin{equation}  \label{gauge}
\delta_{\epsilon}\phi^i(x)= R^{i}_{\alpha}\epsilon^{\alpha}(x)
\equiv\int d^ny\, R^{i}_{\alpha}(x,y)\epsilon^{\alpha}(y),
\end{equation}
a functional $W$ depending on the original fields $\phi ^i$ and the
ghosts $C ^{\alpha}$, together with their associated antifields $\phi ^*_i$ 
and $C ^*_{\alpha}$. This functional possesses
the following properties :

\begin{enumerate}
\item[$\bullet$]  $W$ is bosonic and has ghost number zero,

\item[$\bullet$]  $W$ starts like {\small $W=I_{0}+\phi _{i}^{*}R_{\alpha
}^{i}C^{\alpha }+\frac{1}{2}C_{\gamma }^{*}C_{\alpha \beta }^{\gamma
}C^{\alpha }C^{\beta }+\frac{1}{4}\phi _{i}^{*}\phi _{j}^{*}M_{\alpha \beta
}^{ij}C^{\alpha }C^{\beta }+\mbox{``more''}$} where ``more'' contains
at least three ghosts,

\item[$\bullet$]  $W$ fulfills the ``master equation''\newline
\begin{equation}
(W,W)=0  \label{master_equation}.
\end{equation}
\end{enumerate}

The ``antibracket'' $(.\;,.\;)$ makes the fields $\phi^i$ and the ghosts $%
C^{\alpha}$ canonically conjugate respectively to the antifields $\phi^*_i$
and the antighosts $C^*_{\alpha}$. It is defined by : 
\begin{equation}  \label{antibracket}
(A,B)= \frac{\delta^R A}{\delta \phi^i}\frac{\delta^L B}{\delta \phi^*_i} -%
\frac{\delta^R A}{\delta \phi^*_i}\frac{\delta^L B}{\delta \phi^i} +\frac{%
\delta^R A}{\delta C^{\alpha}}\frac{\delta^L B}{\delta C^*_{\alpha}} -\frac{%
\delta^R A}{\delta C^*_{\alpha}}\frac{\delta^L B}{\delta C^{\alpha}}
\end{equation}
where the superscript $R$ (resp. $L$) denotes a right (resp. left) derivative.
The antibracket satisfies the graded Jacobi identity and increases the ghost
 number by one unit, {\it{i.e.}} $gh((A,B))=gh(A)+gh(B)+1$.

The master equation (\ref{master_equation}) is fulfilled as a consequence of
the Noether identities 
\begin{equation}  \label{Noether}
\frac{\delta I_0}{\delta \phi^i} R^{i}_{\a} =0
\end{equation}
and the gauge algebra : 
\begin{equation}  \label{algebra}
R^{j}_{\alpha}(\phi)\frac{\delta R^{i}_{\beta}(\phi)}{\delta \phi^j}
-(-)^{\epsilon_{\epsilon_{\alpha}}\epsilon_{\eta_{\beta}}}R^{j}_{\beta}
(\phi)\frac{\delta R^{i}_{\alpha}(\phi)}{\delta\phi^j} =
C^{\gamma}_{\alpha\beta}(\phi)R^{i}_{\gamma}(\phi)
+M^{ij}_{\alpha\beta}(\phi)\frac{\delta I_0}{\delta \phi^j}, \\
\end{equation}
where $M^{ij}_{\alpha\beta}=-(-)^{\epsilon_{\phi^i}\epsilon_{\phi^j}}
M^{ji}_{\alpha\beta} $.

Conversely, given any $W$ solution of (\ref{master_equation}), one can
recover the gauge-invariant action as the term independent of the ghosts in $%
W$, while the gauge transformations are defined by the terms linear in the
antifields $\phi^*_i$ and the structure functions appearing in the gauge
algebra can be read off from the terms quadratic in the ghosts. The Noether
identities (\ref{Noether}) are fulfilled as a consequence of the master
equation , the gauge algebra (\ref{algebra}) and of all the higher order
identities that one can derive from them. In other words,
there is complete equivalence between gauge invariance of $I_0$ and the
existence of a solution $W$ of the master equation. For this reason, one can
reformulate the problem of consistently introducing interactions for a gauge
theory as that of deforming $W$ while maintaining the master equation (\ref
{master_equation}).

\subsection{Perturbation of the master equation}

Let $W_0$ be the solution of the master equation for the original theory, 
\begin{equation}
W_0=I_0+\phi^*_iR^{i}_{\alpha}C^{\alpha}, \mbox{\hspace{1cm}}(W_0,W_0)=0.
\end{equation}
Because the gauge transformations are abelian, there is no further term in $%
W_0$ . Let W be the solution of the master equation for the searched-for
interacting theory, in the deformation parameter $g$, we have  
\begin{eqnarray}
W_0&=&I_0+\phi^*_iR^i_{\alpha}C^{\alpha}+O(C^2), \\
I &=& I_0+gI_1+g^2I_2+\cdots, \\
(W,W)&=&0.
\end{eqnarray}
and $W$ exits if and only if $I$ is a consistent deformation of $I_0$.
Let us expand $W$ and the master equation for $W$ in powers of the
deformation parameters $g$. With $W=W_0+gW_1+g^2W_2+O(g^3) $, the equation $%
(W,W)=0$ yields, up to order $g^2$ : 
\begin{eqnarray}
O(g^0)& :\mbox{\hspace{0.5cm}}& (W_0,W_0)=0,  \label{33} \\
O(g^1)& :\mbox{\hspace{0.5cm}}&(W_0,W_1)=0,  \label{34} \\
O(g^2)& :\mbox{\hspace{0.5cm}}&(W_0,W_2)=-\frac{1}{2}(W_1,W_1).  \label{35}
\end{eqnarray}
The first equation is fulfilled by assumption since the starting point
defines a consistent theory. $W_0$, the solution of the master equation for
the free theory is in fact the generator of the BRST differential $s$ : 
\begin{equation}  \label{36}
sA=(W_0,A) \mbox{ for a functional } A.
\end{equation}
the nilpotency $s^2=0$ follows from the master equation for $W_0$ and the
graded Jacobi identity for the antibracket. Thus equation (\ref{34})
expresses that $W_1$ is a BRST-cocycle, {\it{i.e.}} that it is ``closed'' 
under $s$ : 
\begin{equation}  \label{37}
sW_1=0.
\end{equation}
Trivial interactions generated by field-redefinitions that reduce to the
identity at order $g^0$ precisely correspond to cohomologically trivial
solution of (\ref{37}), i.e., correspond to ``exact'' $A$ (also called
``coboundaries'') of the form $A=sB$ for some $B$.

We thus come to the conclusion that the non-trivial consistent interactions
are characterized to first order in $g$ by the \emph{cohomological group} 
\footnote{%
We recall that, given some nilpotent $s$, $s^2=0$, $H(s)$ denotes the
equivalence classes of ``closed'' $A$'s, modulo ``exact'' ones, i.e. the
solution of $sA=0$, modulo the equivalence relation $A^{\prime}=A+sB$.} $H(s)
$ at ghost number zero. 

Because the equation $s \int a =0$ is equivalent to $sa+dm=0$ (where $d$
denotes the Cartan's exterior differential) for some $m$,
and $\int a = s \int b$ is equivalent to $a=sb+dn$ for some $n$, one denotes
 the corresponding cohomological group by $H^{0,n}(s\vert d)$
 \footnote{%
More generally, we shall use the notation $H^{i,p}_j$ to denote a
cohomological group for $p$-forms having a fixed ghost number $i$, and a
fixed ``antifield'' number $j$. If we indicate only one superscript, it will
always refer to the form degree $p$.}
, where  $a,b,m,n$ are {\it{local forms}}, that is, differential forms with
 local functions as coefficients. 
{\it{Local functions}} depend polynomially on the fields - including the 
ghosts and the antifields - and their derivatives up to a finite order, in 
such a way that we work with functions over a finite-dimensional vectorial 
space, the so-called jet space.

Once a first-order deformation is given, one must investigate whether it can
be extended to higher orders. It is a direct consequence of the graded
Jacobi identity for the antibracket that $(W_1,W_1)$ is BRST-closed. However, 
it may not 
be BRST-exact (in the space of local functionals). In this
case, the first-order deformation $W_1$ is obstructed at second-order, so,
it is not a good starting point. If, on the other hand, $(W_1,W_1)$ is
BRST-exact, then a solution $W_2$ to (\ref{35}), which may be rewritten 
\begin{equation}
sW_2=-\frac{1}{2}(W_1,W_1)
\end{equation}
exits. Since $(W_1,W_1)$ has ghost number one (because the antibracket 
increases
the ghost number by one unit), we see that obstructions to continuing a
given, first order consistent interaction are measured by the cohomological
group $H^{1,n}(s|d)$. Furthermore, the ambiguity in $W_2$ (when it exists)
is a solution of the homogeneous equation $sW_2=0$.

The same pattern is found at higher orders : obstructions to the existence of 
$W_k$ are elements of $H^{1,n}(s|d)$, while the ambiguities in $W_k$ (when
it exists)are elements of appropriate quotient spaces of $H^{0,n}(s|d)$.%
\newline

To compute $H^0(s|d)$, we need the following cohomological groups :
$H(\gamma)$, $H(\gamma|d)$, $H(\delta)$, $H(\delta|d)$ and $%
H^{inv}(\delta|d) $ as is expressed by the following theorem (see \cite
{antifield_I}) 
\begin{theorem}[Cohomology of  $s$ modulo $d$]\label{tsd}
\begin{eqnarray}
\mbox{ (i) } H^k(s|d) &\simeq& H_{-k}(\delta|d) \mbox{ with } k<0 \label{e20}\\
\mbox{(ii) } H^k(s|d) &\simeq& H^k(\gamma|d,H_0(\delta)) \mbox{ with  } k 
\geq 0 
\end{eqnarray}
  $k$ is the ghost number, except on  the right-hand side of 
(\ref{e20}), where it stands for the antighost number.
\end{theorem}

\section{Cohomology of the gravitino}
\setcounter{equation}{0}  
\setcounter{theorem}{0}  
\setcounter{lemma}{0} 
\label{cohom32} 

Following the formalism described in the previous section,
we now search for all possible consistent deformations of the lagrangian $%
\mathcal{L}_{\frac{3}{2}}$.

\subsection{$H(\gamma)$ for the gravitino.}

We isolate the contractible pairs $\partial_{\mu_1\cdots\mu_p}C=\gamma(%
\partial_{\mu_1\cdots\mu_{p-1}}\psi_{\mu_p})$ with respect to the
differential $\gamma$ as in \cite{DVHTV}. This shows that all derivatives of
the ghost are $\gamma$-exact and thus are trivial in $H(\gamma)$.
Furthermore, the only gauge-invariant objects constructed out of the fields
$\psi_{\m}$ and their derivatives are the antisymmetrized first order 
derivatives of the
fields : $\psi_{\mu\alpha }\equiv\partial_{[\alpha}\psi_{\mu]}$ and their
subsequent derivatives. Thus, 
\begin{theorem} [Cohomology of $\gamma$ : $H(\gamma)$]
Let  $a$ be a local function\footnote{
The notation $f([\phi])$ means that $f$ is a function of $\phi$ and its 
subsequent derivatives up to a finite order.
},
$
\gamma(a)=0 \Rightarrow a=f([\psi_{\mu\alpha }],[{\psi}^*_{\mu }],[C^*],C)+
\gamma b$, with some local function  $b$.
\end{theorem}

As the derivatives of the ghost $C$ are $\gamma $-trivial, one can
express all the invariant local functions as 
\begin{equation}
\alpha _{J}([\psi _{\mu \alpha }],[{\psi }_{\mu }^{*}],[C^{*}])\omega ^{J}(C)
\end{equation}
The $\alpha _{J}$ are local functions of $\psi _{\mu \alpha }$, ${\psi }_{\mu
}^{*}$ , $C^{*}$ and their subsequent derivatives together with forms $%
dx^{\mu }$.\newline
The $\omega ^{J}$ are monomials which constitute a basis for the algebra
generated by the ghost $C$. In opposition to
the case of gravitons (see \cite{graviton} section 3), this algebra is not
finite-dimensional, because in our case the ghost $C$ is bosonic.

The $\alpha_J$ which contain a finite number of derivatives are polynomials
at each antighost numbers, because $\psi_{\mu\alpha}$ and $C^*$ being
fermionic variables, appear only a finite number of time when one fixes the
number of derivatives. The occurrence of ${\psi}^*_{\mu }$ is limited by the
fixation of the antighost number. In the sequel, we will only consider this
case and the $\alpha_J([\psi_{\mu\alpha }],[{\psi}^*_{\mu }],[C^*])$ will be
called \emph{invariant polynomials}. In antifield number zero, the invariant
polynomials are functions of the $\psi_{\mu\alpha }$ and their subsequent
derivatives .\newline

In what follows, demonstrations of  theorems follow the same lines as
those of similar theorems in \cite{graviton} with some little adaptations.

\subsection{Invariant cohomology of $d$ : $H^{inv}(d)$}

We will need to compute the cohomology of the Cartan's differential $d$ in
the space of invariant polynomials : 
\begin{theorem}[Cohomology of  $d$ in the space of invariant polynomials]
\label{Th24}

In antighost number  strictly greater than 0 and in form degree less than 
the spacetime dimension, the  cohomology of  $d$ is trivial 
in the space of invariant polynomials.
 In other words, if  $a$ is an invariant polynomial, the equation 
$da=0$ ( with antigh$(a)>0$)
implies a solution  $a=db$ with some invariant polynomial  $b$.
\end{theorem}

The demonstration follows the pattern shown in \cite{DVHTV}. 
Namely, split $d$ as $d=d_{1}+d_{0}$,
where $d_{1}$ acts only on the antifields ($C^{*}$, $\psi ^{*\mu }$) and $%
d_{0}$ acts only on the field $\psi _{\mu \nu }$. By the Poincar\'{e} Lemma
( see section 4 of \cite{rept_cohomology}) $d_{1}$ has no cohomology in form
degree less than the space-time dimension (and in antifield number strictly
greater than 0) because there is no relation among the derivatives of the
antifields. By contrast, $d_{0}$ has some cohomology in the space of
polynomials in $\psi _{\mu \nu }$ (see the previous section). From the
triviality of the cohomology of $d_{1}$, one easily gets $d\alpha
=0\Rightarrow d\beta +u$, where $\beta $ is an invariant polynomial, and
where $u$ is an invariant polynomial that does not involve the antifields.
However, since $antigh(\alpha )>0$, $u$ must vanish. \mbox{}\nolinebreak
\hfill \rule{2mm}{2mm}\medbreak 

This theorem has the following consequence : 
\begin{theorem}\label{Th25}
If $antigh(a)>0$  then $\gamma a + db=0$
is equivalent (up to trivial redefinition) to 
$
\gamma a=0
$
\end{theorem}
The proof is similar to the demonstration in section $A.1$ of \cite
{graviton}. The difference lies in the fact that in our case, the first
order derivatives of the ghosts are $\gamma$-exact. Hence we conclude easily
without introducing new differential to analyse the descent equations. %
\mbox{}\nolinebreak\hfill\rule{2mm}{2mm}\medbreak

\subsection{Cohomology of $\delta$ modulo $d$ : $H(\delta|d)$}

The cohomological group $H(\delta|d)$ is related to conservations laws 
and more 
generally to the so-called characteristic cohomology through the isomorphism 
$H^{n-p}_{char}(d)\simeq H^n_p(\delta|d)$ \cite{DVHTV}(for a review of this
 aspect of the 
differential $\delta$, see  section 6 of \cite{rept_cohomology}).

$H_l(\delta|d)$ is trivial in the space of local forms with pureghost
number $l>0$ (see \cite{antifield_I}), we will then, in the sequel of this 
subsection,
consider only local forms with $puregh=0$.

Because the theory is linear and irreducible we have the 
(see \cite{antifield_I}) 

\begin{theorem}
[Cohomology of $\delta$ modulo $d$ : $H^n_p(\delta|d), p>2$]
\label{Th27}
The cohomological groups  $H^n_p(\delta|d)$ vanish in antifield number 
strictly 
greater than 2,
\begin{equation}
H^n_p(\delta|d)=0 \mbox{ with } p> 2.
\end{equation}
\end{theorem} 

This theorem means that all the conservation laws involving antisymmetric
objects of rank strictly greater than 2 are trivial (see section 6.2 of 
\cite
{rept_cohomology}).\\

In antifield number two, the cohomology is given by the following theorem :

\begin{theorem}[Cohomology of $\delta$ modulo $d$ : $H^n_2(\delta|d)$]
\label{th28}
A complete set of representatives of 
$H^n_2(\delta|d)$ is given by the antighost $C^*$ :
$
\delta a^n_2+da^{n-1}_1=0 \Rightarrow 
a^n_2=\lambda_rC^{*r}\,d^nx+\delta b^n_3+db^{n-1}_2
$;
where  $\lambda_r$ are constants.
\end{theorem}

The proof follows similar lines as for Theorem 4.2 of \cite{graviton}. Once
again the demonstration is easier than in the case of graviton, thanks to
the first degree in the derivatives of the spinor theory.\newline

Let $a$ be a solution of the cocycle condition for $H^n_2(\delta|d)$,
written in dual notations, $\delta a+\partial_{\mu}V^{\mu}=0$. Without loss
of generality, one can assume that $a$ is linear in the undifferentiated
antifields ( integrations by parts leaves one in the same cohomological
class of $H^n_2(\delta|d)$). Thus 
\begin{equation}  \label{e24}
a=\lambda_rC^{*r}+b
\end{equation}
where $b$ is quadratic in the antifields $\psi^{*\mu}$ and their derivatives,
and where $\lambda_i$ are functions of $\psi_{\mu}$ and their derivatives.
Because $\delta\mu\approx 0$\footnote{$\approx$ means equal modulo equations
of motion.}, the equation $\delta a+\partial_{\mu}V^{\mu}=0$ implies $%
\partial_{\mu}\lambda_r{\psi}^{*r\mu}\approx 0$. By the linear dependency of
the $\psi^{*\mu}$, we conclude that $\partial_{\mu}\lambda_r\approx 0$.
Thanks to the isomorphism $H^0_0(d|\delta)/R\simeq H^n_n(\delta|d)$ (see
Theorem 6.2 of \cite{rept_cohomology}) and the previous theorem $%
H^n_n(\delta|d)=0$ ($n>2$), we conclude that $\lambda_r\approx l_r$, where
 $l_r$
are constants. Substituting this expression into (\ref{e24}) and
noting that the term proportional to the equations of motion can be absorbed
through a redefinition of $b$, one gets $a=l_rC^r+b^{\prime}$ up to trivial
terms). Now $l_rC^r$ is a solution of $\delta a+\partial_{\mu}V^{\mu}=0$ by
itself. This means that $b^{\prime}$, which is quadratic in the $\psi^{*\mu}$
and their derivatives, must be a $\delta$-cocycle modulo $d$, and hence
trivial (see Theorem 11.2 of \cite{antifield_I}). \mbox{}\nolinebreak\hfill%
\rule{2mm}{2mm}\medbreak

\subsection{Invariant cohomology of $\delta$ modulo $d$ : $H^{inv}(\delta|d)$}

In the space of invariant polynomials, we have :

\begin{theorem}[Invariant cohomology of $\delta$ modulo $d$ : 
$H^{inv}(\delta|d)$]\label{th29}
Let  $a$ be an invariant   polynomial. If $a$ is  $\delta$-trivial 
modulo $d$ in the space of all polynomials  (including non-invariant ones), 
$a=\delta b +dc$, then  $a$ is also  $\delta$-trivial modulo $ d $ in the 
space of invariant polynomials , that is to say, one can takes 
$b$ and  $c$ as  invariant polynomials.
\end{theorem}

The proof is similar to the one of Theorem 4.1 of \cite{graviton}. It follows
descent equations techniques as (A.9) of \cite{antifield_II}. Once again, in
our case we conclude easily, thanks to the first order of the spinor theory. %
\mbox{}\nolinebreak \hfill \rule{2mm}{2mm}\medbreak 

\section{Consistent deformations for the gravitino}

\label{cons32}

\subsection{Cohomology of $s$ modulo $d$}

A cocycle of $H(s|d)$ is a solution of 
\begin{equation}
sa + db=0.
\end{equation}

We expand $a$ and $b$ as a series indexed by the antifield number (for this
section, see section A.3 of \cite{graviton} and \cite{antifield_I}) : 
\begin{eqnarray}
a &=&a_{0}+a_{1}+\cdots +a_{k}, \\
b &=&b_{0}+b_{1}+\cdots +b_{l}.
\end{eqnarray}

The two series stop at some finite antifield number, because the first-order
deformation of the lagrangian is assumed to have a finite derivative order 
\cite{antifield_I}. As $H^n_k(\delta|d)$ is trivial for $k\geq3$ (see
theorems \ref{tsd} and \ref{Th27}), we can stop with $a_2$ : $a=a_0+a_1+a_2$
and $b=b_0+b_1$ (see \cite{antifield_I}). Using $s=\gamma + \delta$, we have
the following ``descent equations'' : 
\begin{eqnarray}
\delta a_1 +\gamma a_0 + db_0 &=& 0,  \nonumber \\
\delta a_2 +\gamma a_1 + db_1 &=& 0,  \nonumber \\
\gamma a_{2}=0.
\end{eqnarray}
Furthermore, the term $a_2$ can be taken to contain only no-trivial terms of 
$H_2^n(\delta |d)$.

Now we have all the tools to compute $H^{0,n}(s|d)$. Note that 
\noindent $a_0$ is the deformation of the lagrangian, $a_1$ gives the
deformation of the gauge transformations and $a_2$ contains information on
the gauge algebra.

We start with $a_2$ and then ``lift'' it two times in order to find $a_1$ 
and $a_0$.
\subsection{The algebra of gauge transformations remains abelian after
consistent deformation}
The general solution of $\gamma a_2=0$ is modulo trivial terms : 
\begin{equation}
a_2=\alpha_J\omega^J
\end{equation}
with $\alpha_J$ invariant polynomials.
$a_2$ can be taken to contain non-trivial terms of $H^n_2(\delta |d)$, because
if $t_2=\alpha^{\prime}_J\omega^J$ is such that $\delta t_2 +dc=0$, then we
get $st_2 +dc=0$ (because $s=\gamma+\delta$ and $\gamma(\alpha^{\prime}_J%
\omega^J)=0$, hence $t_2$ can be taken to zero.
Thus, $\alpha_J$ are of the type ${\lambda}_r C^{*r}$, where $\lambda_r$ are 
constants. But, as $\omega^J$ contains two ghosts, $a_2$ must vanish
because there is no Lorentz invariant expression built out of 3 spinors.
We conclude that $a_2=0$, which means that there is no deformation of the 
algebra of
gauge transformation : the gauge algebra remains abelian under deformation.

\subsection{Deformations of the gauge transformation}

Now, the equations are : 
\begin{eqnarray}
\delta a_1 + \gamma a_0 &=& db_0, \nonumber \\
\gamma a_1 &=& 0.
\label{4.31}
\end{eqnarray}

Equation (\ref{4.31}) implies $a_{1}=\alpha _{J}\omega ^{J}$. As the ghost 
number vanish and the
antighost is 1, we have to build $a_{1}$ out of the ghost $C$ and an
antifield $\psi ^{*\mu }$. To respect the Lorentz invariance, one must
contract the space-time indices of $\psi _{\mu }$ with a derivatives $%
\partial ^{\mu }$ or gamma matrices. By adding a total derivative 
if necessary, we can put all the derivatives on the ghost $C$, 
and then the expression becomes $\gamma$-trivial.
Hence, without imposing any restriction on the number of derivatives, 
the unique possibility is then
\footnote{We use $W_0=I_0-\bar{\psi}^{*\m}\partial_{\m}C$}
 \begin{equation}
-im\bar{\psi} ^{*\mu }\gamma _{\mu }{C}.
\end{equation}
This correspond to an abelian gauge transformation as anticipated by the
vanishing $a_{2}$. 
\begin{equation}
\delta _{\epsilon }\psi _{\mu }=\mathcal{D}_{\mu }\epsilon =\partial _{\mu
}\epsilon +im\gamma _{\mu }\epsilon.
\end{equation}
Note that $\mathcal{D}_{\mu}$ is not a derivative (it does not respect the
Leibnitz rule, this operator is known in the full nonlinear supergravity 
$D=4$ $N=1$ (see \cite{broken,cosmological_constant}). 
$m$ is the deformation parameter.

It is interesting to recall that, in the case of the pure spin-2 studied 
in \cite{graviton},
there was no control on the number of derivatives for the equation
$\g a_1=0$. Here, the pure spin-$\frac{3}{2}$ case is much more constrained.
\subsection{The only first-order consistent deformation of the lagrangian 
is a mass term}

With $a_1=-im\bar{\psi}_{\m}^*\g^{\m}C$, we compute
\begin{equation}
\d a_1=im\partial_{\nu}\bar{\psi}_{\rho}
\gamma^{\mu\nu\rho} \gamma_{\mu}{C}.
\end{equation}
As $\gamma^{\mu\nu\rho}\gamma_{\mu}=2\gamma^{\nu\rho}$, the equation above
  gives 
\begin{equation}
\d a_1=2im\bar{C}\gamma^{\nu\rho}\partial_{\nu}\psi_{\rho},
\end{equation}
which can written as 
\begin{equation}
\d a_1= \partial_{\nu}(2im\bar{C}\gamma^{\nu\rho}\psi_{\rho})
 -2im\gamma(\bar{\psi_{\nu}}%
)\gamma^{\nu\rho}\psi_{\rho}.
\end{equation}
Hence, 
\begin{equation}
\d a_1=\partial_{\nu}(2im\bar{C}\gamma^{\nu\rho}\psi_{\rho}) 
-im\gamma(\bar{\psi}%
_{\nu}\gamma^{\nu\rho}\psi_{\rho})
\end{equation}
from which we get : 
\begin{eqnarray}
a_0 &=& im\bar{\psi}_{\nu}\gamma^{\nu\rho}\psi_{\rho} \\
&=& -\frac{i}{2}m\bar{\psi}_{\mu}\gamma^{\mu\nu\rho}\gamma_{\nu}\psi_{\rho}.
\end{eqnarray}
This is a mass term
which is obstructed at second order \cite{Bautieretal} (this can be seen
from the expression $(W_{1},W_{1})$, which is not $s$-exact and is hence an
obstruction).
This is in agreement with general belief that mass term and gauge invariance 
are incompatible.

The total lagrangian is 
\begin{eqnarray}
\mathcal{L}_{\frac{3}{2}}+\mathcal{L}_{M} &=&-\frac{1}{2}\bar{\psi}_{\mu
}\gamma ^{\mu \nu \rho }\partial _{\nu }\psi _{\rho }-\frac{i}{2}m\bar{\psi}%
_{\mu }\gamma ^{\mu \nu \rho }\gamma _{\nu }\psi _{\rho } \\
&=&-\frac{1}{2}\bar{\psi}_{\mu }\gamma ^{\mu \nu \rho }\mathcal{D}_{\nu }\psi
_{\rho }
\end{eqnarray}
and is invariant under the gauge transformation $\delta _{\epsilon }\psi
_{\mu }=\mathcal{D}_{\mu }\epsilon $ to first order in the coupling constant 
$m$. The gravitino-mass term was already known in the full nonlinear theory
of N=1 D=4 supergravity \cite{cosmological_constant,broken}.

\section{Consistent interactions between a spin-2 and a spin-3/2 field}

\setcounter{equation}{0}  
\setcounter{theorem}{0}  
\setcounter{lemma}{0}  

In this section, we will search all the consistent deformations of 
$\mathcal{L}_2+%
\mathcal{L}_{\frac{3}{2}}$ to first order in the deformation parameters
under the assumptions made in the introduction, namely
locality, Poincar\'e invariance, conservation of the number of gauge 
symmetries and of the number of derivatives acting on the fields.

As we found in the previous section, the only deformations $\mathcal{L}%
_{M}$ involve only the gravitino $\psi_{\mu}$. The
deformations involving only the graviton were computed in \cite{graviton}%
\footnote{%
This was $\mathcal{L}_E+\mathcal{L}_C$, where $\mathcal{L}_C$ is the
cosmological term with no derivative of the field and $\mathcal{L}_E$ is
the term of the Einstein-Hilbert action of third order in the field $%
h_{\mu\nu}$, this term has two derivatives.}.
All we have to do is to find
interactions terms involving both the gravitino $\psi_{\mu}$ and the
graviton $h_{\mu\nu}$ simultaneously.

\subsection{Cohomology of $\mathcal{L}_2+\mathcal{L}_{\frac{3}{2}}$}

As the theory is the sum of two free lagrangians, we have 
\begin{equation}
H(\gamma)=H(\gamma_2)\otimes H(\gamma_{\frac{3}{2}}).
\end{equation}
Using result of our section \ref{cohom32} and of \cite{graviton} on the
cohomology of $\gamma $, if $a$ is a local function of the fields
(including all the spectrum of the Batalin-Vilkovisky formalism)
, then\footnote{$K$ stands for the linearized Riemann
tensor} : 
\begin{equation}
a=\alpha _{JJ^{\prime }}([K],[h^{*}],[\xi ^{*}],[\partial _{[\mu }\psi _{\nu
]}],[\psi _{*}],[C_{*}])\omega ^{J^{\prime }}(C)\omega ^{J}(\xi _{\mu
},\partial _{[\mu }\xi _{\nu ]}).
\end{equation}

The following theorems are quite direct by using the similar theorem for the
two free lagrangians :

\begin{theorem}[Invariant cohomology of $d$ : $H^{inv}(d)$]
In form degree less than the spacetime dimension and in antifield number 
strictly greater than zero, the cohomology of $d$ in the space of invariant 
polynomials is trivial.
\end{theorem}

\begin{theorem}[Cohomology of $\delta$ modulo $d$ : $ H^n_p(\delta |d)=0 
\;\;\mbox{ 
for  }\;p>2.$]
$H^n_p(\delta |d)$ is trivial in antifield number strictly greater than 2,
\begin{equation}
H^n_p(\delta |d)=0 \;\;\mbox{ for  }\;p>2.
\end{equation}
\end{theorem}

\begin{theorem}[Cohomology of $\delta$ modulo $d$ : $H^n_2(\delta|d)$]
The cohomology of $H^n_2(\delta| d)$ is generated 
by the antighosts $C^{*r},\xi^*_{\mu}$ :
\begin{equation}
\delta a^n_2 + \delta a_1^{n-1}=0
\Rightarrow a^n_2=({\lambda_r}C^{*r} + f_{\mu}\xi^{*\mu})
dx^0dx^1\cdots dx^{n-1}+\delta b^n_3+db^{n-1}_2
\end{equation}
where $\lambda_r$ and $f_{\mu}$ are constants.
\end{theorem}

\begin{theorem}[Invariant cohomology of $\delta$ modulo 
$d$ : $H^{inv}(\delta|d)$]
Let $a$ be  $\delta$-trivial modulo $d$ and $\gamma$-invariant, 
$a=\delta b+dc$, 
then $a$ is $\delta$-trivial in the space of invariant polynomials. 
That is to say, one can choose $b$ and $c$  in the space of invariant 
polynomials.
\end{theorem}

\subsection*{Cohomology of $s$ modulo $d$}

Let $a$ be  ghost number zero solution of 
\begin{equation}  \label{327}
sa + db=0.
\end{equation}
For the same reason as in the case of the gravitino, we can expend $a$ and $%
b$ according to the antighost number as : 
\begin{eqnarray}
a&=&a_0+a_1+a_2, \\
b&=&b_0+b_1,
\end{eqnarray}

with $s=\gamma+\delta$ we get the descent equations : 
\begin{eqnarray}
\delta a_1 + \gamma a_0 = db_0, \\
\delta a_2 + \gamma a_1 = db_1, \\
\gamma a_2 =0.
\end{eqnarray}
Once more, we recall that $a_0$ is the deformation of the lagrangian, $a_1$
gives the deformation of the gauge transformations and $a_2$ contains
information on the gauge algebra.

\subsection{Deformations of the algebra of gauge symmetries}

The general solution of $\gamma a_2=0$ which involve mixed terms 
(that is with spin-2 and spin-$\frac{3}{2}$) and respect Poincar\'e 
invariance is, modulo
trivial terms, \footnote{%
As in the case of gravitino we can ignore trivial terms of 
$H^n_2(\delta|d)$} : 
\begin{equation}
a_2=\alpha\frac{1}{4}\xi^{*\alpha}\bar{C}\gamma_{\alpha}C+\frac{\beta}{4}
\partial_{[\alpha}\xi_{\beta]}\bar{C^*}\gamma^{\alpha\beta}C
+\frac{\lambda}{4}\xi_{\alpha}\bar{C}^*\gamma^{\alpha}C,
\end{equation}
where the factors in front of the coefficients are chosen for 
further convenience.

\subsection{Deformation of the gauge symmetries}

In this subsection we study whether one can lift $a_2$ to a certain $a_1$ 
by looking at solution of $\delta a_2 + \gamma a_1 =db_1$.

\subsubsection{Lift of $a_2=\frac{1}{4}\xi_{\alpha}^*\bar{C}\gamma^{\alpha}C$
                to $a_1$}

Let $a_2=\frac{1}{4}\xi_{\alpha}^*\bar{C}\gamma^{\alpha}C$, then 
\begin{eqnarray}
\delta a_2 &=& -\frac{1}{2}\partial_{\beta}h^{*\alpha\beta}\bar{C}%
\gamma_{\alpha}C \\
&=& \partial_{\beta}(-\frac{1}{2}h^{*\alpha\beta}\bar{C}\gamma_{\alpha}C)
+h^{*\alpha\beta}(\partial_{\beta}\bar{C})\gamma_{\alpha}C \\
&=&\partial_{\beta}(-\frac{1}{2}h^{*\alpha\beta}\bar{C}\gamma_{\alpha}C)
-\gamma(h^{*\alpha\beta}\bar{\psi}_{\beta}\gamma_{\alpha}C)
\end{eqnarray}

from which we get 
\begin{equation}
a_1 = h^{*\alpha\beta}\bar{\psi}_{\alpha}\gamma_{\beta}C.
\end{equation}
modulo $\gamma$-exact terms .

\subsubsection{Lift of $a_2=\partial_{[\alpha}\xi_{\beta]}\bar{C_*}%
\gamma^{\alpha\beta}C$ to $a_1$}

Let $a_2 = \partial_{[\alpha}\xi_{\beta]}\bar{C_*}\gamma^{\alpha\beta}C$, as
the matrix $\gamma^{\alpha\beta}$ is antisymmetric with respect to $%
\alpha\beta$ we write $a_2 = \partial_{\alpha}\xi_{\beta}\bar{C_*}%
\gamma^{\alpha\beta}C$.

\begin{eqnarray}
\delta a_2 &=& \delta(\partial_{\alpha}\xi_{\beta}\bar{C}_*\gamma^{\alpha%
\beta}C) \\
&=& \partial_{\alpha}\xi_{\beta}\partial_{\lambda}\bar{\psi}^{\lambda}_*
\gamma^{\alpha\beta}C \\
&=& \partial_{\lambda}(\partial_{\alpha}\xi_{\beta}\bar{\psi}^{\lambda}_*
\gamma^{\alpha\beta}C ) -\partial_{\lambda\alpha}\xi_{\beta}\bar{\psi}%
^{\lambda}_* \gamma^{\alpha\beta}C -\partial_{\alpha}\xi_{\beta}\bar{\psi}%
^{\lambda}_* \gamma^{\alpha\beta}\partial_{\lambda}C.
\end{eqnarray}
As 
\begin{eqnarray}
\partial_{\lambda\alpha}\xi_{\beta}\bar{\psi}^{\lambda}_*
\gamma^{\alpha\beta}C &=& \gamma [ \frac{1}{2} ( \partial_{\lambda}
h_{\alpha\beta}+\partial_{\alpha} h_{\beta\lambda} -\partial_{\beta}
h_{\lambda\alpha} ) \bar{\psi}^{\lambda}_*\gamma^{\alpha\beta}C ] \\
&=&\gamma(\partial_{\alpha}h_{\lambda\beta} \bar{\psi}^{\lambda}_*\gamma^{%
\alpha\beta}C)
\end{eqnarray}
and 
\begin{equation}
-\partial_{\alpha}\xi_{\beta}\bar{\psi}^{\lambda}_*
\gamma^{\alpha\beta}\partial_{\lambda}C = \gamma (
\partial_{\alpha}\xi_{\beta}\bar{\psi}^{\lambda}_*
\gamma^{\alpha\beta}\psi_{\lambda} )
\end{equation}
we obtain : 
\begin{eqnarray}
a_1 &=& \partial_{\alpha}h_{\lambda\beta} \bar{\psi}^{\lambda}_*\gamma^{%
\alpha\beta}C -\partial_{\alpha}\xi_{\beta}\bar{\psi}^{\lambda}_*
\gamma^{\alpha\beta}\psi_{\lambda} \\
&=& \partial_{\alpha}h_{\lambda\beta} \bar{C}\gamma^{\alpha\beta}\psi^{%
\lambda}_* -\partial_{\alpha}\xi_{\beta}\bar{\psi}_{\lambda}
\gamma^{\alpha\beta}\psi^{\lambda}_*.
\end{eqnarray}

\subsubsection{Obstruction for $a_2=\xi_{\alpha}\bar{C}^*\gamma^{\alpha}C$}

Let $a_2=\xi_{\alpha}\bar{C}_*\gamma^{\alpha}C$,
\begin{eqnarray}
\delta a_2 &=& -\xi_{\alpha}\partial_{\beta}\bar{\psi}_*^{\beta}\gamma^{%
\alpha}C \\
&=& \partial_{\beta}(-\xi_{\alpha}\bar{\psi}_*^{\beta}\gamma^{\alpha}C)
+\partial_{\beta}\xi_{\alpha}\bar{\psi}_*^{\beta}\gamma^{\alpha}C
+\xi_{\alpha}\bar{\psi}_*^{\beta}\gamma^{\alpha}\partial_{\beta}C \\
&=& \partial_{\beta}(-\xi_{\alpha}\bar{\psi}_*^{\beta}\gamma^{\alpha}C)
+\partial_{[\beta}\xi_{\alpha]}\bar{\psi}_*^{\beta}\gamma^{\alpha}C
+\partial_{(\beta}\xi_{\alpha)}\bar{\psi}_*^{\beta}\gamma^{\alpha}C 
\nonumber \\
& & +\xi_{\alpha}\bar{\psi}_*^{\beta}\gamma^{\alpha}\partial_{\beta}C.
\end{eqnarray}
The last two terms are $\gamma$-exact : 
\begin{eqnarray}
\xi_{\alpha}\bar{\psi}_*^{\beta}\gamma^{\alpha}\partial_{\beta}C
&=&-\gamma(\xi_{\alpha}\bar{\psi}_*^{\beta}\gamma^{\alpha}\psi_{\beta} ) \\
\partial_{(\beta}\xi_{\alpha)}\bar{\psi}_*^{\beta}\gamma^{\alpha}C
&=&\gamma(h_{\beta\alpha}\bar{\psi}_*^{\beta}\gamma^{\alpha}C).
\end{eqnarray}
The term $\partial _{[\beta }\xi _{\alpha ]}\bar{\psi}_{*}^{\beta }\gamma
^{\alpha }C$ is clearly non-trivial in $H(\gamma |d)$. That can be seen as
follows : Suppose that 
\begin{equation}
\partial _{[\beta }\xi _{\alpha ]}\bar{\psi}_{*}^{\beta }\gamma ^{\alpha
}C=\gamma u+\partial _{\mu }v^{\mu }
\end{equation}
by taking the Euler-Lagrange derivative with respect to $\bar{\psi}_{*}^{\mu
}$ of both side of the previous relation, one gets that $\frac{\delta ^{L}}{%
\delta \bar{\psi}_{*}^{\mu }}\partial _{[\beta }\xi _{\alpha ]}\bar{\psi}%
_{*}^{\beta }\gamma ^{\alpha }C$ is $\gamma $-exact, because $\gamma $
commutes with $\frac{\delta }{\delta \psi _{*}^{\mu }}$ and $\frac{\delta }{%
\delta \psi _{*}^{\mu }}\partial _{\mu }v^{\mu }=0$.\newline
In fact, one gets : $\frac{\delta }{\delta \bar{\psi}_{*}^{\mu }}\partial
_{[\beta }\xi _{\alpha ]}\bar{\psi}_{*}^{\beta }\gamma ^{\alpha }C=-\partial
_{[\mu }\xi _{\alpha ]}\gamma ^{\alpha }C$ which clearly not $\gamma $-exact
as term of the form $\partial _{[\mu }\xi _{\alpha ]}$ are not $\gamma _{2}$%
-exact (see section 3 of \cite{graviton}).

The most general mixed term for $a_1$ is thus,

\begin{equation}
a_1=\alpha h^{*\alpha\beta}\bar{\psi}_{\beta}\gamma_{\alpha}C +\frac{\beta}{4}
(\partial_{\alpha}h_{\lambda\beta} \bar{C}\gamma^{\alpha\beta}\psi^{%
\lambda}_* -\partial_{\alpha}\xi_{\beta}\bar{\psi}_{\lambda}
\gamma^{\alpha\beta}\psi^{\lambda}_*)+\gamma e.
\label{5.72}
\end{equation}
Note that lifting $a_2$ to $a_1$ has reduced the number of free parameters
for the interaction part of $W$ from 3 ($\alpha$, $\beta$, $\lambda$) to 2 (%
$\alpha$, $\beta$) thanks to the obstruction of $a_2=\xi_{\alpha}\bar{C}%
^*\gamma^{\alpha}C$.

\subsection{Deformation of the lagrangian}

$\mathcal{L}_{Int}=a_0$ is the solution of $\delta a_1 + \gamma a_0 =db_0$.
\\
Starting with 
\begin{eqnarray}
a_1&=&\a(h^{*\a\b}\bar{\psi}_{\b}\g_{\a}C)+
\nonumber \\
&+&\frac{\b}{4}(\pa_{\a}h_{\l\b}\bar{\psi}^{*\l}\g^{\a\b}C-
\pa_{\a}\x_{\b}\bar{\psi}^{*\l}\g^{\a\b}\psi_{\l})
\end{eqnarray}
we get
\begin{eqnarray}
\d a_1 &=&
\alpha(\Box h^{\alpha\beta} +\partial^{\alpha\beta}h
-\partial^{\alpha}\partial_{\rho}h^{\rho\beta}
-\partial^{\beta}\partial_{\rho}h^{\rho\alpha}+  
\nonumber \\
&+&\eta^{\alpha\beta}\partial_{\rho\lambda}h^{\rho\lambda}
-\eta^{\alpha\beta}\Box h)\bar{\psi}_{\beta}\gamma_{\alpha}C+
\nonumber \\
&+&
-\frac{\b}{4}(\underbrace{ \pa_{\a}h_{\l\b} \pa_{\m}\bar{\psi}_{\n} 
\g^{\m\n\l} \g^{\a\b}
 C }_{I}+
\pa_{\a}\x_{\b}\pa_{\m}\bar{\psi}_{\n}\g^{\m\n\l}\g^{\a\b}\psi_{\l}).
\end{eqnarray}
Up to total derivatives, the expression $I$ writes
\be
I=-[\pa_{\m\a}h_{\b\l}\bar{\psi}_{\n}\g^{\m\n\l}\g^{\a\b}C+
\pa_{\a}h_{\b\l}\bar{\psi}_{\n}\g^{\m\n\l}\g^{\a\b}(\g \psi_{\m})].
\ee
Using 
\bqn
\g^{\m\n\l}\g^{\a\b}&=&\stackrel{(3)}{[\g^{\m\n\l}\g^{\a\b}]}+
\stackrel{(1)}{[\g^{\m\n\l}\g^{\a\b}]}
\nonumber \\
&=&
\left[ \right.
\eta^{\alpha\lambda}\gamma^{\beta\mu\nu}
+\eta^{\alpha\mu}\gamma^{\beta\nu\lambda}
+\eta^{\alpha\nu}\gamma^{\beta\lambda\mu}+
 \nonumber \\
&+& (-)\eta^{\beta\lambda}\gamma^{\alpha\mu\nu}
-\eta^{\beta\mu}\gamma^{\alpha\nu\lambda}
-\eta^{\beta\nu}\gamma^{\alpha\lambda\mu}  \left. \right]+
\nonumber \\
&+& \left[ \right. \eta^{\beta\lambda}\eta^{\alpha\mu}\gamma^{\nu}
+\eta^{\beta\mu}\eta^{\alpha\nu}\gamma^{\lambda}
+\eta^{\beta\nu}\eta^{\alpha\lambda}\gamma^{\mu}  \nonumber \\
& & -\eta^{\beta\lambda}\eta^{\alpha\nu}\gamma^{\mu}
-\eta^{\beta\nu}\eta^{\alpha\mu}\gamma^{\lambda}
-\eta^{\beta\mu}\eta^{\alpha\lambda}\gamma^{\nu} \left. \right]
\eqn
we have also
\bqn
I&=&\pa_{\m}V^{\m}+[(\Box h^{\alpha\beta} +\partial^{\alpha\beta}h
-\partial^{\alpha}\partial_{\rho}h^{\rho\beta}
-\partial^{\beta}\partial_{\rho}h^{\rho\alpha}+  
\nonumber \\
&+&\eta^{\alpha\beta}\partial_{\rho\lambda}h^{\rho\lambda}
-\eta^{\alpha\beta}\Box h)\bar{\psi}_{\beta}\gamma_{\alpha}C]-
\pa_{\m\a}h_{\b\l}\bar{\psi}_{\n}
[\stackrel{(3)}{\g^{\m\n\l}\g^{\a\b}}]C+
\nonumber \\
&+&\frac{1}{2}\g \left( \right.
\pa_{\a}h_{\b\l}\bar{\psi}_{\n}[\stackrel{(1)}{\g^{\m\n\l}\g^{\a\b}}]
\psi_{\m} \left. \right)+
\nonumber \\
&-&\pa_{\a}\x_{\b}\pa_{\m}\bar{\psi}_{\n}[\stackrel{(1)}{\g^{\m\n\l}\g^{\a\b}}]
\psi_{\l}-\pa_{\m\a}h_{\b\l}\bar{\psi}_{\n}
[\stackrel{(3)}{\g^{\m\n\l}\g^{\a\b}}]\pa_{\m}C.  
\eqn
Also, as $\bar{\psi}_{\n}[\stackrel{(3)}{\g^{\m\n\l}\g^{\a\b}}]\psi_{\m}=0$,
we obtain the following result for $\d a_1$ :
\bqn
\d a_1 &=&\pa_{\m}V^{\m}+\g \left[ \frac{\b}{8}\bar{\psi}_{\m}\g^{\m\n\l}
\stackrel{(1)}{\o}_{\n}\psi_{\l}\right]+
\nonumber \\
&+&(\a-\frac{\b}{4})\left[ \right.(\Box h^{\alpha\beta} 
+\partial^{\alpha\beta}h
-\partial^{\alpha}\partial_{\rho}h^{\rho\beta}
-\partial^{\beta}\partial_{\rho}h^{\rho\alpha}+  
\nonumber \\
&+&\eta^{\alpha\beta}\partial_{\rho\lambda}h^{\rho\lambda}
-\eta^{\alpha\beta}\Box h)\bar{\psi}_{\beta}\gamma_{\alpha}C\left. \right]+
\nonumber \\
&+&-\frac{\b}{4}  \pa_{\a}h_{\b\l}\pa_{\m}\bar{\psi}_{\n}
[\stackrel{(3)}{\g^{\m\n\l}\g^{\a\b}}]C-\frac{\b}{4}\pa_{\a}\x_{\b}\pa_{\m}
\bar{\psi}_{\n}[\stackrel{(3)}{\g^{\m\n\l}\g^{\a\b}}]\psi_{\l},
\eqn
where $\stackrel{(1)}{\o}_{\l}= \pa_{\a}h_{\b\l}\g^{\b\a}$ is the spin 
connection at first order in the field $h_{\m\n}$\footnote{Note the slight 
abuse of notation : by 
$\o_{\m}$ we really mean $\o_{\m}^{~ab}\g_{ab}$, with $\o_{\m}^{~ab}$ the
spin connection.}.
The previous expression has the advantage to give us already a part of $a_0$,
the deformation of the lagrangian :
\be
a_0= \ldots-\frac{\b}{8}\bar{\psi}_{\m}\g^{\m\n\l}\stackrel{(1)}
{\o}_{\n}\psi_{\l}.
\ee
We may also get rid of the terms beginning with 
$\Box h^{\alpha\beta}\bar{\psi}_{\beta}\gamma_{\alpha}C$ by choosing 
$\a=\frac{\b}{4}$.
Once those simplifications are done and the first piece of $a_0$ is discarded, 
we are left with two terms in $\d a_1$, that we have to express as 
$\g$-exact terms plus total derivatives. These two terms contain three gamma
matrices explicitely. 
In order to express them has $\g$-exact terms modulo total derivatives, we give
a basis of $a_0$-terms which correspond to the two remaining pieces 
of $\d a_1$.
This basis reads 
\beq
\{Q_\Delta\}_{0\leq\Delta\leq 3}&=&\{ h\bar{\psi}_{\a}\g^{\a\m\n}\pa_{\m}
\psi_{\n},  h_{\a\b}\bar{\psi}_{\m}\g^{\a\m\n}\pa^{\b}\psi_{\n},
\nonumber \\ 
 &&h_{\a\b}\bar{\psi}_{\m}\g^{\a\m\n}\pa_{\n}\psi^{\b},
 h_{\a\b}\bar{\psi}^{\b}\g^{\a\m\n}\pa_{\m}\psi_{\n} \}.
\eqn
We then compute $\g (\a^{\Delta}Q_\Delta)$, and try to match this with
$\d a_1^{remaining} + \pa_{\m}V^{\m}$. This gives a system of equations
for the coefficients $\a^{\Delta}$ which is solved for the following values, 
following the same order as
for the $Q_{\Delta }$ : 
\begin{equation}
\alpha ^{\Delta }=\{-\frac{\b}{4},-\frac{\b}{4},\frac{\b}{4},\frac{\b}{4} \}.
\end{equation}
Actually the system is not completely solved, namely there remains a term 
in  $\frac{-1}{4}\g (\b Q_0+ \b Q_1 -\b Q_2-\b Q_3)$ which does not match 
anything in the remaining terms of $\d a_1$. The complete equation 
$\d a_1 + \g a_0$ is a total derivative modulo the following term :
\be
-2\b \x^{\b}\pa_{[\a}\bar{\psi}_{\b ]}\g^{\a\m\n}\pa_{[\m}{\psi}_{\n ]}.
\ee
It obviously belongs to $H(\g)$. However, a rapid check immediately tells us 
that this term can be absorbed
in $a_1$, because is $\d$-exact.
The new part that $a_1$ acquires is
\be
a_1 \rightarrow a_1 +2\b \x^{\l}\bar{\psi}^{*\a}\pa_{[\a}{\psi}_{\l ]}.
\ee
It is amusing to see what this term corresponds to.
The equation $\d a_2 + \g a_1=dc_1$ gave us the $a_1$ in (\ref{5.72}), 
modulo a 
solution $\bar{a}$ of the homogeneous equation $\g \bar{a}_1+dc_1=0$.
This last equation, being of strictly positive antighost number, is equivalent
to $\g \bar{a}_1=0$, as theorem \ref{Th25} learns us.
Now, asking that the deformed lagrangian does not bring more than one 
derivative on the $\psi_{\m}$, the only candidate $a_1$ belonging to $H(\g)$ 
is 
precisely $\x^{\b}\bar{\psi}^{*\a}\pa_{[\a}{\psi}_{\l ]}$, if we don't use
any gamma matrices, and demanding Lorentz invariance.
The final $a_0$ is finally
\bqn
a_0 &=&-\frac{\b}{2} \left[ \right.
\bar{\psi}_{\a}\g^{\a\m\n}\frac{1}{4}\stackrel{(1)}{\o}_{\m}\psi_{\n}+
\nonumber \\
&+&(-) \bar{\psi}_{\a}\g^{\a\m\n}\frac{h^{\b}_{~\m}}{2}\pa_{\b}\psi_{\n}-
\bar{\psi}_{\a}\g^{\a\m\n}\frac{h^{\b}_{~\n}}{2}\pa_{\m}\psi_{\b}-
\bar{\psi}_{\b}\frac{h^{\b}_{~\a}}{2}\g^{\a\m\n}\pa_{\m}\psi_{\n}+
\nonumber \\
&+& \frac{h}{2}\bar{\psi}_{\a}\g^{\a\m\n}\pa_{\m}\psi_{\n}\left. \right ].
\label{vertex}
\eqn
This is indeed the right cubic vertex of $N=1,D=4$ supergravity.
To convince the reader we recall that the complete lagrangian is
\be
\cl\propto
e \bar{\psi}_{\a}\g^{abc}e_a^{~\a}e_b^{~\m}e_c^{~\n}D_{\m}\psi_{\n},
\ee
while at first order
\bqn
g_{\m\n}&\equiv&e^a_{~\m}e_{a\n}=\h_{\m\n}+ h_{\m\n},
\nonumber \\
e^a_{~\m}&=&\d^a_{~\m}+ \frac{1}{2}h^a_{~\m},
\nonumber \\
e_a^{~\m}&=&\d_a^{~\m}- \frac{1}{2}h_{~a}^{\m},
\nonumber \\
e&\equiv&\sqrt{-g}=1+\frac{1}{2}h_{\m\n}\h^{\m\n}\equiv 1+\frac{1}{2}h.
\eqn
The complete lagrangian contains the covariant derivative
\be
D_{\m}\psi_{\n}=\pa_{\m}\psi_{\n}+\frac{1}{4}\o_{\m}\psi_{\n}.
\ee
Taking into account the results for the gravitino alone, the non-trivial
 expression of $a_1$ is~: 
\begin{equation}
a_1=-\frac{\b}{4} h^{*\alpha\beta}\bar{\psi}_{\beta}\gamma_{\alpha}C 
+\frac{\beta}{4}
(\partial_{\alpha}h_{\lambda\beta} \bar{C}\gamma^{\alpha\beta}\psi^{%
\lambda}_* -\partial_{\alpha}\xi_{\beta}\bar{\psi}_{\lambda}
\gamma^{\alpha\beta}\psi^{\lambda}_*)
-im\bar{\psi}_{\m}^*\g^{\m}C
+2\b \bar{\psi}^{*\a}\x^{\l}\pa_{[\a}{\psi}_{\l ]}.
\end{equation}

The last term $+2\b \bar{\psi}^{*\a}\x^{\l}\pa_{[\a}{\psi}_{\l ]}$ gives
 us the Lie derivative of $\psi_{\a}$ along the
vector $\x^{\l}$ : after partial integration of
$\b\bar{\psi}^{*\a}\x^{\l}\pa_{\a}\psi_{\l}$
we get 
$ -\b \pa_{\a}\bar{\psi}^{*\a}\x^{\l}\psi_{\l}$ 
which is $\d$-exact and is thus absorbed through a trivial redefinition 
$a_2\rightarrow a_2 + \g[\b \bar{C}^*\x^{\l}\psi_{\l}] $, 
plus $-\b \bar{\psi}^{*\a}\pa_{\a}\x^{\l}\psi_{\l}$ which, combined 
with $-\b\bar{\psi}^{*\a}\x^{\l}\pa_{\l}\psi_{\a}$ is the Lie derivative
of the covector $\psi_{\a}$.\\
{\it{The existence of an interaction vertex automatically implies the 
Lie derivative of the gravitino as a gauge symmetry of the theory.}}

\section{Consistency to second order and uniqueness of the deformation}
\setcounter{equation}{0}  
\setcounter{theorem}{0}  
\setcounter{lemma}{0} 

Putting the results of the two previous sections and of \cite{graviton}
together, we find that the most general consistent deformation of the
lagrangian $\mathcal{L}_{2}+\mathcal{L}_{\frac{3}{2}}$, which is local,
respect Poincar\'{e} invariance and the number of gauge transformations and
derivative of each field is : 
\begin{equation}  \label{deform_lagrangian}
\mathcal{L}=\mathcal{L}_{2}+\mathcal{L}_{\frac{3}{2}}+g\mathcal{L}%
_{E}+\alpha \mathcal{L}_{int}+m\mathcal{L}_{M}+\Lambda \mathcal{L}_{C}.
\end{equation}
The master equation at second order in the couplings constants is : $(W_1,
W_1)=-2sW_2$. This implies that $(W_1, W_1)$ must be BRST-trivial. By
looking at the terms of maximal antighost number $g\xi^{*\alpha}
\xi^{\beta}\partial_{[\beta}\xi_{\alpha]} +\alpha(\xi^{*\alpha}\bar{C}%
\gamma_{\alpha}C+\partial^{\alpha} \xi^{\beta}\bar{C}^*
\gamma_{\alpha\beta}C)$ this gives the relations 
\begin{equation}
4\alpha^2-g\alpha = 0.
\end{equation}
The solution $\alpha =0$ is not consider because it forbids interactions, we
then have 
\begin{equation}
g=4\alpha.
\end{equation}
This relationship had been obtained through the related Noether method
 (see \cite{Deser:zb}).
On the other hand, if we want $(W_1, W_1)$ to be BRST-trivial, we must have
an other relation : 
\begin{equation}
\a\Lambda-3m^2=0.
\end{equation}
Hence, consistency to second order leaves us with only two free coupling
constants : $g$ and $m$.

\subsection{Analysis of the deformation}

We will now show that the deformed theory corresponds to $D=4$ $N=1$
supergravity.

\subsubsection{Analysis of the lagrangian}

First note that the deformed lagrangian is now : 
\begin{equation}
\mathcal{L}=\mathcal{L}_{2}+\mathcal{L}_{\frac{3}{2}}+g\mathcal{L}_{E}+ 4g%
\mathcal{L}_{int}+m\mathcal{L}_{M}+\frac{12m^2}{g}\mathcal{L}_{c}.
\end{equation}
This corresponds to the linearized lagrangian of $D=4$ $N=1$ supergravity
with a cosmological constant $\Lambda =\frac{3m^{2}}{\a}=\frac{12m^2}{g}$.

\subsubsection{Analysis of the gauge symmetries}

As $a_1$ is related to the gauge transformations of the fields. 
The most general non-trivial $a_1$ is : 
\begin{equation}
a_1=-\frac{g}{4} h^{*\alpha\beta}\bar{\psi}_{\beta}\gamma_{\alpha}C 
+\frac{g}{4}
(\partial_{\alpha}h_{\lambda\beta} \bar{C}\gamma^{\alpha\beta}\psi^{%
\lambda}_* -\partial_{\alpha}\xi_{\beta}\bar{\psi}_{\lambda}
\gamma^{\alpha\beta}\psi^{\lambda}_*)
+2g \x^{\l}\bar{\psi}^{*\a}\pa_{[\a}{\psi}_{\l ]}
-im\bar{\psi}^{*\m}\g_{\m}C.
\end{equation}
This corresponds to the gauge transformations for the gravitino :
\begin{equation}
\delta _{\epsilon ,\z }\psi _{\lambda }=
\pa_{\l}\eps+
\frac{g}{4} \pa_{\a}h_{\b\l}\g^{\b\a}C
+\frac{g}{4}\pa_{\a}\z_{\b}\g^{\a\b}\psi_{\l}
+g(\z^{\a}\pa_{\a}\psi_{\l}+\pa_{\l}\z^{\a}\psi_{\a})+im\g_{\l}\eps
\end{equation}
where $\epsilon $ is a spinor and $\eta $ a 4-vector. The first two terms on
the right-hand side correspond to the linearized covariant derivatives : 
\begin{equation}
D_{\lambda }=\partial _{\lambda }+\frac{1}{4}\stackrel{(1)}{\omega} _{\lambda }
\end{equation}
where $\stackrel{(1)}{\omega} _{\lambda }=-\partial _{\mu }h_{\lambda \beta
}\gamma ^{\mu \beta }$. The third term is a linearized Lorentz transformation.
The fourth term is the Lie derivative of the covector $\psi_{\a}$ along the
diffeomorphism vector. The last one is the mass term.
For the graviton, we get : 
\begin{equation}
\delta_{\epsilon}h_{\mu\nu}=-\frac{g}{2}(\bar{\psi}_{\mu}\gamma_{\nu}%
\epsilon+\bar{\psi}_{\nu}\gamma_{\mu}\epsilon)
\end{equation}
which is the linearized supergravity gauge symmetry for the graviton.

\subsubsection{Analysis of the algebra of gauge symmetries}

Commutators of gauge transformations are related to $a_2$.

With 
\begin{equation}
a_2=\frac{g}{4}(\frac{1}{2}\xi^{*\alpha}\bar{C}\gamma_{\alpha}C-
\partial_{[\alpha}\xi_{\beta]}\bar{C^*}\gamma^{\alpha\beta}C)
\end{equation}
we get :

\begin{equation}
[\epsilon_1,\epsilon_2]\eta_{\alpha}=\frac{1}{2}\bar{\epsilon}%
_1\gamma_{\alpha}\epsilon_2
\end{equation}

and

\begin{equation}
[\eta_{\alpha},\epsilon]\epsilon^{\prime}=-\partial_{[\alpha}\eta_{\beta]}%
\gamma^{\alpha\beta}\epsilon^{\prime}
\end{equation}

This is indeed the $N=1$ $D=4$ supersymmetric algebra.
Hence we have proved that our deformation corresponds to $D=4$ $N=1$
linearized supergravity with a possible cosmological term . As we know that 
this latter is consistent to all orders, we have proved that
supersymmetry (through supergravity) is the only way to introduce consistent
interactions between a massless spin 2 and a massless spin 3/2-field under
the assumptions stated in the introduction.

\section*{Acknowledgements}  
We are very grateful to Marc Henneaux for suggesting the
problem.
This work is partially supported by the ``Actions de  
Recherche Concert{\'e}es" of the ``Direction de la Recherche  
Scientifique - Communaut{\'e} Fran{\c c}aise de Belgique", by  
IISN - Belgium (convention 4.4505.86)
and by the European Commission RTN programme
HPRN-CT-00131
in which N. B. is associated to K. U. Leuven.  
M.E. is grateful to Christiane Schomblond for fruitful discussions.

\end{document}